\documentclass[pra,twocolumn,amsmath,amssymb,superscriptaddress]{revtex4-1}

\usepackage{epsfig,amsmath}
\usepackage{subfigure}
\usepackage{graphicx}% Include figure files
\usepackage{dcolumn}% Align table columns on decimal point
\usepackage{stmaryrd}
\usepackage{mathrsfs}
\usepackage{pifont}
\usepackage{amsthm}
\usepackage{amssymb}
\usepackage{bm}
\usepackage{latexsym}
\usepackage[colorlinks=true,linkcolor=blue,citecolor=blue]{hyperref}
\usepackage{color}
\usepackage{epstopdf}

\begin{document}

\title{Simultaneous cooling by measuring one ancillary system}

\author{Jia-shun Yan}
\affiliation{Department of Physics, Zhejiang University, Hangzhou 310027, Zhejiang, China}

\author{Jun Jing}
\email{Email address: jingjun@zju.edu.cn}
\affiliation{Department of Physics, Zhejiang University, Hangzhou 310027, Zhejiang, China}

\date{\today}

\begin{abstract}
We present a simultaneous-cooling protocol for a double-resonator system via projective measurements on an ancillary $V$-type qutrit. Through repeated measurements on the ground state of the ancillary system, the two resonators can be cooled down to their respective ground states from thermal states. With respect to the measurement-based cooling, an optimized measurement-interval $\tau_{\rm opt}$ is analytically obtained for the first time, which is inversely proportional to the collective thermal Rabi frequency $\Omega_{\rm th}$ as a function of the resonators' average population of the last round. Under about only $20$ optimized measurements, the average population of the target resonators can be reduced by $6$ orders in magnitude. Our simultaneous or collective cooling protocol is scalable to the systems with more numbers of resonators and robust to the fluctuation in the resonator frequency.
\end{abstract}

\maketitle

\section{Introduction}

Preparing quantum systems into their ground-state is a defining feature of quantum physics, leading to state manipulation, storage, and conversion that find no counterpart in classical physics. Beyond its importance at a fundamental level~\cite{OpenQuantumSystem,QuantumNoise,QuantumOptics,LaserCooling}, the ground-state cooling is also a crucial component for many modern quantum technologies~\cite{MechanicalOscillatorCooling,MechanicalSidebandCooling}, such as continuous-variable quantum computations ~\cite{QuantumComputation,ComputationRobustness,YouSuperconductingCircuits}, boson sampling~\cite{BosonSampling,PhononArithmetic}, and ultrahigh-precision measurements~\cite{ForceMeasureByOscillator1,ForceMeasureByOscillator2}. Great efforts have been made to develop the ground-state cooling techniques, e.g., the sideband cooling in ion-trap systems with dissipative channels and the dilution refrigerator in superconducting circuits.

Unlike the paradigms dependent on cryogenic temperature, the nondeterministic ground-state cooling of resonator modes via measuring an ancillary system with limited degrees of freedom under control has been theoretically proposed~\cite{Purification,OneModeCooling} and experimentally verified~\cite{ExpOneModeCooling}. The composite system in this approach undergoes a joint unitary evolution for a constant or random interval of time before a projective measurement is taken on the ground-state of the ancillary system, which is bonded to the ground-state of the resonator. The evolution-measurement procedure is repeated, conditioning on the events for which the outcome is found to be successful. Otherwise, the sample in the ensemble is discarded to yield a nonunit successful probability~\cite{ExpOneModeCooling}. Measurement-based cooling is interesting on its own for simulating the quantum Zeno effect by postselection. It has been generalized to various scenarios~\cite{OneWayComputer,ExpOneWayComputer,MeasureEntanglement}, e.g., cooling a nonlinear mechanical resonator~\cite{NonLinearCooling} and cooling by one-shot-measurement~\cite{OneShotCooling}.

In practical applications including coherent-mode mixing~\cite{DynamicsOfTwomodes}, quantum interface~\cite{MechanicalInterface}, indirect state transfer assisted by a controllable ancillary system~\cite{ThreeResonatorCircuit}, and hybrid-system entanglement~\cite{MechanicalInterface}, the to-be-cooled target system has more than one motional mode~\cite{TwoModeStateTransfer}. Efficient ground-state cooling for a larger number of modes becomes even more desired for high-fidelity quantum manipulations~\cite{EITCoolingTwoMode} and interesting to many-body physics. In contrast to the existing techniques for multi-mode cooling, such as sideband cooling in optomechanical systems~\cite{TwoModeSidebandCooling,TwomodeSimultaneousCooling}, electromagnetically-induced-transparency (EIT) cooling in ion crystals~\cite{EITCoolingTwoMode}, and phonon cooling by the three-wave parametric interaction~\cite{ThreeWaveCooling}, we find that the measurement-based cooling exhibits both high-efficiency and scalability without extra efforts to designing energy-dissipative channels.

In this work, we propose a simultaneous or collective cooling-by-measurement protocol for a double-resonator system. The two resonators (the target system) are respectively coupled to the level transitions of $|g\rangle\leftrightarrow|e\rangle$ and $|g\rangle\leftrightarrow|f\rangle$ in a $V$-type three-level atom (the ancillary qutrit). Here $|g\rangle$, $|e\rangle$, and $|f\rangle$ are ground and two excited levels, respectively. It is shown that rounds of projective measurements with equal-time-spacing on $|g\rangle$ are capable to simultaneously cool down both resonators to their own ground states, irrespective to the initial temperatures. For the first time, we obtain an analytical expression to estimate the optimal interval with respect to the cooling performance, which is inversely proportional to the collective thermal Rabi frequency as a function of the resonators' average population of the last round. It allows us to iterate the interval between two consecutive measurements, leading to an ultrafast cooling protocol by unequal-time-spacing measurements. In addition, both protocol and the optimal interval of projective measurement adapt to the situation with more resonators coupled to one multilevel atom, in which the $k$th mode is coupled to the transition between the ground state and the $k$th excited level of the ancillary system. All resonators are found to be cooled down to their own ground states via periodically measuring the common ground state of the ancillary system.

The structure of this work is as follows. In Sec.~\ref{HamAndCoolOperater}, we introduce our model for cooling two resonator modes coupled to a $V$-type three-level system and briefly describe the general framework of the measurement-based cooling. A double-mode-cooling coefficient is obtained to describe the population-reduction ratios for each Fock state occupied by both modes. In Sec.~\ref{TwoModeCoolingSec}, we present the effect of the measurement interval on the average-population reduction, by which we derive an analytical expression for an optimized interval. Then we provide the optimized cooling performance under both equal-time-spacing and unequal-time-spacing strategies of measurement. In Sec.~\ref{MultimodeCoolingSec}, our protocol is extended to the multi-mode situation. In Sec.~\ref{Discussion}, we discuss the cooling performance out of the multi-photon resonant condition and compare our protocol with the other cooling methods. We also comment the analogy between cooling-by-measurement and state-preparation by signal herald. Finally, the results are summarized in Sec.~\ref{Conclusion}.

\section{Cooling model and coefficient}\label{HamAndCoolOperater}

\begin{figure}[htbp]
\centering
\includegraphics[width=0.95\linewidth]{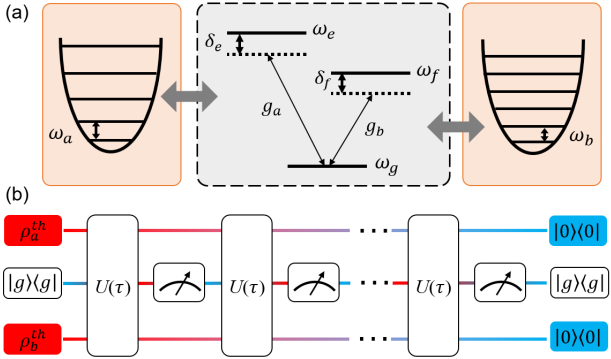}
\caption{(a) Diagram of our double-mode-cooling model employing a V-type three-level atom (the ancillary system) and two non-degenerate resonator modes (the target system), whose eigen-frequencies are respectively $\omega_a$ and $\omega_b$. (b) Circuit model for our double-mode-cooling protocol. Starting from the thermal states, mode-$a$ (the upper line) and mode-$b$ (the bottom line) are simultaneously cooled down to their own ground states, under a number of rounds of free-evolution and instantaneous measurement performed on the ground state of the ancillary atom (the middle line).}\label{model}
\end{figure}

Our cooling protocol is based on a composite system shown in Fig.~\ref{model}(a), consisted of a $V$-type three-level atom (qutrit) and two harmonic oscillators (resonators). The three-level system with a $V$ configuration can be realized in both natural atoms~\cite{VtypeThreeLevel} and artificial ones~\cite{YouSuperconductingCircuits}; and the oscillators could be the resonators in superconducting circuit~\cite{OneModeCooling}, the mirror's displacement induced by radiation~\cite{TwoModeSidebandCooling}, or the motional modes in trapped ion crystals~\cite{EITCoolingTwoMode}. The qutrit (the ancillary system) has two excited levels $|e\rangle$ and $|f\rangle$ and a ground state $|g\rangle$, and the two resonators (the target system) are labeled by $a$ and $b$, respectively. Then the overall Hamiltonian of the composite system reads ($\hbar\equiv 1$)
\begin{equation}\label{Ham}
\begin{aligned}
H=&\omega_g|g\rangle\langle g|+\omega_e|e\rangle\langle e|+\omega_f|f\rangle\langle f|+\omega_aa^\dagger a+\omega_b b^\dagger b\\ &+g_a\left(a^\dagger\sigma_{eg}^-+a\sigma_{eg}^+\right)+g_b\left(b^\dagger\sigma_{fg}^-+b\sigma_{fg}^+\right),
\end{aligned}
\end{equation}
where $\omega_i$ is the frequency of the qutrit level $|i\rangle, i=g,e,f$, $a$ and $b$ ($a^\dagger$ and $b^\dagger$) are annihilation (creation) operators of the two resonator modes with frequencies $\omega_a$ and $\omega_b$, respectively. For simplicity and with no loss of generality, the ground-state energy is set as $\omega_g=0$. It is reasonable to assume that two modes are non-degenerate and they are respectively near-resonant with the transitions $|g\rangle\leftrightarrow|e\rangle$ and $|g\rangle\leftrightarrow|f\rangle$. Otherwise there is no motivation to design a double-mode cooling protocol. The coupling between ancillary system and resonators is of the Jaynes-Cummings (JC) type and the coupling strengths with mode-$a$ and mode-$b$ are respectively $g_a$ and $g_b$. The transition operators of the three-level system are denoted by $\sigma^+_{ij}=|i\rangle\langle j|$ and $\sigma^-_{ij}=|j\rangle\langle i|$. In the rotating frame with respect to $H_0=\omega_a(a^\dagger a+|e\rangle\langle e|)+\omega_b(b^\dagger b+|f\rangle\langle f|)$, the full Hamiltonian reads
\begin{equation}\label{InteractionHam}
\begin{aligned}
H_I=&\delta_e|e\rangle\langle e|+\delta_f|f\rangle\langle f|+g_a\left(a^\dagger\sigma_{eg}^-+a\sigma_{eg}^+\right) \\
&+g_b\left(b^\dagger\sigma_{fg}^-+b\sigma_{fg}^+\right),
\end{aligned}
\end{equation}
where $\delta_e\equiv\omega_e-\omega_a$ and $\delta_f\equiv\omega_f-\omega_b$ are two detunings between atomic levels and resonator modes. Following a general setting of measurement-based cooling, our protocol concatenates the free unitary evolutions under the Hamiltonian $H_I$ and the instantaneous measurements, as demonstrated by the circuit model in Fig.~\ref{model}(b). The two resonators are initially prepared in their own thermal-equilibrium states with arbitrary temperatures (see the first and the third lines of the circuit model), and the ancillary qutrit starts from the ground state $|g\rangle$. The overall initial state then reads
\begin{equation}\label{InitialState}
	\rho(0)=|g\rangle\langle g|\otimes\rho_a^{\rm th}\otimes\rho_b^{\rm th},
\end{equation}
where $\rho_a^{\rm th}$ and $\rho_b^{\rm th}$ represent the thermal states of both modes. In our protocol, the three-level system is measured by a projection operator $M_g=|g\rangle\langle g|$ acted on its ground state immediately after each period of evolution by $U(\tau)=\exp(-iH_I\tau)$, where $\tau$ is the period of time to be optimized. If the measurement outcome turns out that the ancillary qutrit is excited to either high-energy level, then the system sample in the ensemble is discarded and the whole process restarts. If the qutrit is found to be at the ground state $|g\rangle$, then the whole process continues to the next round. After $N$ rounds of evolutions and measurements with equal-time spacings, the state of two resonators takes the form of
\begin{equation}\label{rhoab}
\rho_{ab}(N\tau)=\frac{V_g(\tau)^N\rho_a^{\rm th}\rho_b^{\rm th}V_g^\dagger(\tau)^N}{P_g(N)}
\end{equation}
where $V_g\equiv\langle g|U(\tau)|g\rangle$ is the nonunitary evolution operator living in the space of the target system and
\begin{equation}
P_g(N)\equiv{\rm Tr}\left[V_g(\tau)^N\rho_a^{\rm th}\rho_b^{\rm th}V_g^\dagger(\tau)^N\right]
\end{equation}
represents the successful or survival probability of the ancillary system still in its initial state $|g\rangle$. In the product Fock-state basis $\{|nm\rangle\equiv|n\rangle_a|m\rangle_b\}$ of the two resonators, we have
\begin{equation}\label{Vg}
V_g(\tau)=\sum_{n,m\geq0}\alpha_{nm}(\tau)|nm\rangle\langle nm|,
\end{equation}
where $\alpha_{nm}$ is a double-mode-cooling coefficient for the Fock-state $|nm\rangle$,
\begin{equation}\label{CoolingCoeffExp}
\alpha_{nm}(\tau)=e^{-i\delta\tau/2}\left[\cos{\Omega_{nm}\tau}+i\delta\frac{\sin(\Omega_{nm}\tau)}{2\Omega_{nm}}\right].
\end{equation}
Note to find this compact expression, we have assumed that the frequency detunings between the two resonators and their corresponding level-transitions satisfy the double-photon resonant condition, i.e., $\delta_e=\delta_f=\delta$. And
\begin{equation}\label{TwoModeRabiFrequency}
\Omega_{nm}=\sqrt{g_a^2n+g_b^2m+\delta^2/4}
\end{equation}
is called a double-mode Rabi frequency. Substituting $V_g(\tau)$ in Eq.~(\ref{Vg}) back to Eq.~(\ref{rhoab}), we have
\begin{equation}\label{rhoab2}
\rho_{ab}(N\tau)=\frac{\sum_{nm}|\alpha_{nm}(\tau)|^{2N}p_np_m|nm\rangle\langle nm|}{P_g(N)},
\end{equation}
where $p_{n}$ and $p_{m}$ describe the initial population distributions of mode-$a$ and mode-$b$ over their Fock-state base $|n\rangle$ and $|m\rangle$, respectively, and the survival probability reads
\begin{equation}\label{PgN}
P_g(N)=\sum_{nm}|\alpha_{nm}|^{2N}p_np_m.
\end{equation}
The cooling performance is determined by the modular square of $\alpha_{nm}(\tau)$
\begin{equation}\label{CoolingCoeffSquare}
|\alpha_{nm}(\tau)|^{2}=\frac{\Omega_{nm}^2-(g_a^2n+g_b^2m)\sin^2(\Omega_{nm}\tau)}
{\Omega_{nm}^2}\leq1,
\end{equation}
which serves as the population-reduction ratio for the associated product Fock-state base. After $N$ measurements, $p_np_m$ becomes $p_np_m|\alpha_{nm}|^{2N}/P_g(N)$. The double-mode cooling coefficient $|\alpha_{nm}|^{2}$ in Eq.~(\ref{CoolingCoeffSquare}) takes a similar form as its single-mode counterpart~\cite{OneModeCooling}. If one of the two resonator modes is decoupled from the qutrit, e.g., $g_b=0$ by tuning the excited level $|f\rangle$ to be far off-resonant from mode-$b$, then the double-mode cooling coefficient $\alpha_{nm}(\tau)$ reduces to
\begin{equation}\label{singlemode}
\alpha_n(\tau)=e^{-i\delta\tau/2}[\cos{\Omega_n\tau}+i\delta\sin(\Omega_n\tau)/(2\Omega_n)]
\end{equation}
with $\Omega_n\equiv\sqrt{ng_a^2+\delta^2/4}$, exactly the same as the single-mode one~\cite{OneModeCooling}. More importantly, in the case of the vacuum state of the resonators, i.e., $n=m=0$, both cooling coefficients become unit. $|\alpha_{00}|=1$ means that the populations of both resonators on their ground states are protected by the nonunitary evolution $V_g(\tau)$, at the cost of repeatedly discarding the distribution of the whole system in the manifolds of our double-mode JC model except the ground state. In sharp contrast, when $nm\neq0$, the populations over most of the Fock-state bases $|nm\rangle$ will be gradually cut down with increasing $N$ unless $\sin(\Omega_{nm}\tau)=0$ or $\Omega_{nm}\tau=j\pi$ with integer $j$.

\begin{figure}[htbp]
\centering
\includegraphics[width=0.95\linewidth]{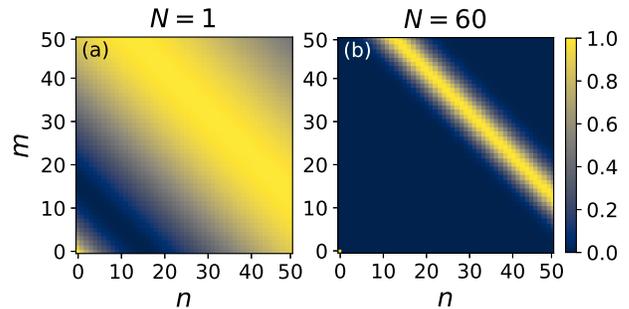}
\caption{Landscape of the double-mode-cooling performance over the space of Fock states $|nm\rangle$ under (a) a single measurement described by $|\alpha_{nm}(\tau)|^2$ and (b) $N=60$ equal-time-spacing measurements described by $|\alpha_{nm}(\tau)|^{2N}$. The frequency detuning between the two resonators and their corresponding level-transitions is fixed as $\delta=0.01\omega_a$. The coupling strengths between the resonator modes and the qutrit are $g_a=g_b=0.04\omega_a$. The measurement interval is $\tau=10/\omega_a$.}\label{CoolingCoefficient}
\end{figure}

We plot the double-mode cooling coefficients distributed over various $n$ and $m$ under a single measurement in Fig.~\ref{CoolingCoefficient}(a) and $60$ measurement in Fig.~\ref{CoolingCoefficient}(b), respectively. The bright or dark areas imply distinct cooling performance over Fock-state bases with particular $n$ and $m$. It indicates that with a sequence of measurements acted on the ancillary system, the average populations for both modes are subject to an overall descent tendency. During the periodical repetition of measurements, the ground state for both modes $|n=0,m=0\rangle$ is always free of population reduction, indicating the possibility of the ground-state cooling. However, approaching the genuine ground-state cooling might be under the restriction of the non-negligible distribution of the initial thermal resonators over certain high-excitation-number Fock states. In Fig.~\ref{CoolingCoefficient}(b), one can find clearly that the Fock-states satisfying $\Omega_{nm}\tau=j\pi$ or $g_a^2n+g_b^2m=j^2\pi^2/\tau^2-\delta^2/4$ are also under protection by $|\alpha_{nm}|=1$. This result limits the range of application in terms of the initial temperature or occupation of the target resonators. To address the cooling-range problem in measurement-based cooling, one might resort to either introducing extra measurement-based cooling by strongly coupling the excited state of the ancillary system to an external level~\cite{ExternalLevelCooling} or optimizing the measurement-interval $\tau$ as we will focus on in the next section.

\section{Optimal measurement-interval and double-mode cooling}\label{TwoModeCoolingSec}

\begin{figure}[htbp]
\centering
\includegraphics[width=0.95\linewidth]{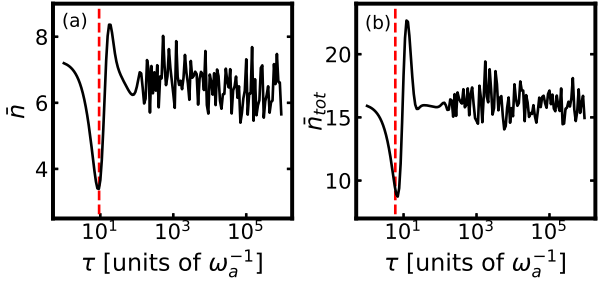}
\caption{The expectation values of the overall populations of the resonator system (the black solid curves) as a function of the measurement-interval $\tau$ for (a) the single mode case obtained by Eq.~(\ref{singlemode}) and (b) the double-mode case obtained by Eq.~(\ref{CoolingCoeffExp}) after a single measurement. The vertical red dashed lines in (a) and (b) indicate the analytical results for the optimized measurement-intervals determined by Eqs.~(\ref{OneModeApproxNum}) and (\ref{ApproxNumTwoMode}), respectively, under $\delta=0$. The parameters for black solid curves are set as $\omega_b=1.2\omega_a$, $T_a=T_b=0.1$ K, $g_a=g_b=0.04\omega_a$, and $\delta=0.01\omega_a$.}\label{OptimalInterval}
\end{figure}

The time-spacing constant $\tau$ between neighbor measurements is a dominant factor in the measurement-based-cooling protocols, determining how fast the projective measurements are performed on the ancillary system and whether or not the target system can be efficiently cooled down. A quantitative and typical observation under a single measurement can be found in Fig.~\ref{OptimalInterval}, where we plot the average populations for a single-mode case and a double-mode case as functions of $\tau$. In particular, we present $\bar{n}$ and $\bar{n}_{\rm tot}$ for mode-$a$ and both modes obtained by the cooling coefficients in Eqs.~(\ref{singlemode}) and (\ref{CoolingCoeffExp}), respectively, under the same setting of the other parameters. The average populations $\bar{n}$ and $\bar{n}_{\rm tot}$ demonstrate similar patterns upon one measurement. It is found that they do not monotonically decline with increasing $\tau$. Instead, the average population declines gradually to the minimal point (about $47\%$ and $55\%$ in magnitude for $\bar{n}$ and $\bar{n}_{\rm tot}$, respectively) at an optimized measurement-interval $\tau_{\rm opt}$, and then rebounds sharply to a peak value that is even larger than the initial population. Finally the dynamics ends up with a high-frequent fluctuation around the initial population due to the thermal state. The resonators could thus be heated up with an inappropriately long measurement-interval. By virtue of the quantum measurement, the populations over almost all the Fock-states of resonators are cut down by the less-than-unit coefficient $|\alpha_n|$ or $|\alpha_{nm}|$, yet the weight of the high-excitation-number Fock-states could be enhanced by the density-matrix renormalization in Eq.~(\ref{rhoab}) or Eq.~(\ref{rhoab2}). It is therefore desired to find an optimized $\tau_{\rm opt}$ to promote cooling rather than heating the resonator systems. While $\tau_{\rm opt}$ is usually attained through numerical optimization or chosen randomly in literature, an analytical expression is interesting for revealing the underlying physics and also instructive for determining the whole evolution time of cooling protocols.

Recalling the single-mode-cooling protocol, its Hamiltonian~\cite{OneModeCooling} can be obtained by ignoring the existence of the level $|f\rangle$ and mode-$b$ in Eq.~(\ref{Ham}),
\begin{equation}
H=\omega_aa^\dagger a+\omega_e|e\rangle\langle e|+g\left(a^\dagger\sigma_{eg}^-+a\sigma_{eg}^+\right).
\end{equation}
Under the resonant condition $\delta=0$, the average population after a single measurement reads
\begin{equation}\label{barn}
\bar{n}=\frac{\sum_n|\alpha_n|^2p_nn}{\sum_n|\alpha_n|^2p_n},
\end{equation}
where $|\alpha_n|^2=\cos^2{\Omega_n\tau}$ is the single-mode-cooling coefficient with the single-mode Rabi frequency $\Omega_n=g\sqrt{n}$ and $p_n$ is the initial population of resonator-$a$ in Fock state $|n\rangle$. It is suggested by Fig.~\ref{OptimalInterval} that the lowest $\bar{n}$ locates nearby the point exhibiting the most dramatically changing in the curve. The idea that this point might be approximated by a singularity leads to the following perturbative analysis over Eq.~(\ref{barn}).

The denominator of Eq.~(\ref{barn}) can be regarded as a summation over $|\alpha_n|^2$ with weight $p_n$, which follows the Maxwell-Boltzmann distribution maximized at the ground state $n=0$ and monotonically declining with increasing the excitation number of Fock states. So that the coefficient $|\alpha_n|^2$ in the denominator can expand around $n=0$ as
\begin{equation}\label{singleOmega}
\cos^2{\Omega_n\tau}=1-g^2n\tau^2+g^4n^2\tau^4/3-\cdots.
\end{equation}
To the order of $\mathcal{O}(\tau^2)$, the average excitation number is approximated as
\begin{equation}\label{OneModeApproxNum}
\bar{n}\approx\frac{\sum_nne^{-\frac{n\omega_a}{k_BT_a}}\cos^2{\Omega_n\tau}}{\sum_n(1-g^2n\tau^2)e^{-\frac{\omega_a n}{k_BT_a}}}=\frac{\sum_nne^{-\frac{n\omega_a}{k_BT_a}}\cos^2{\Omega_n\tau}}{(\bar{n}_{\rm th}+1)(1-\Omega^2_{\rm th}\tau^2)},
\end{equation}
where we have applied the formulas about the geometric series $\sum_{n=0}^{\infty}ne^{-nx}=e^x/(e^x-1)^2$, $T_a$ is the initial temperature of the resonator-$a$, and $\Omega_{\rm th}\equiv g\sqrt{\bar{n}_{\rm th}}$ is defined as the thermal Rabi frequency for the single-mode system with the initial average-population $\bar{n}_{\rm th}=\sum_np_nn$. A singularity of $\tau$ emerges in Eq.~(\ref{OneModeApproxNum}) if
\begin{equation}\label{OptIntervalSingle}
\tau=\tau_{\rm opt}\equiv\frac{1}{\Omega_{\rm th}}.
\end{equation}
This singularity does not really exist, yet providing a regular way to estimating the convergence to the exact numerical solution for the minimal value for $\bar{n}$. That is the vertical red line in Fig.~\ref{OptimalInterval}(a), which is sufficiently close to the optimized value of $\tau$ in the curve. Note the numerator of Eq.~(\ref{barn}) or Eq.~(\ref{OneModeApproxNum}) is a summation over $|\alpha_n|^2$ with weight $p_nn$. The state $|n=[k_BT_a/\omega_a]\rangle$ is thus dominant over the other Fock-states in terms of $p_nn$, where $[\cdot]$ means rounding up or down to an integer. The numerator of Eq.~(\ref{barn}) could then expand around $n=[k_BT_a/\omega_a]$ rather than $n=0$ as in the denominator. Therefore, the ``singularity'' approximation from the denominator could be regarded as the optimal measurement interval.

As for the double-mode cooling in this work, the total average number after one measurement is given by
\begin{equation}\label{barnrot}
\bar{n}_{\rm tot}=\frac{\sum_{nm}|\alpha_{nm}|^2p_np_m(n+m)}{\sum_{nm}|\alpha_{nm}|^2p_np_m}
\end{equation}
under the resonant condition $\delta_e=\delta_f=\delta=0$, where $p_n$ and $p_m$ are the initial populations of resonators $a$ and $b$ in their respective Fock states. In parallel to Eq.~(\ref{singleOmega}), $|\alpha_{nm}|^2$ in the denominator expands around $n=m=0$ as
\begin{equation}
\cos^2\Omega_{nm}\tau=1-\left(g_a^2n+g_b^2m\right)\tau^2+\cdots.
\end{equation}
Then to the order of $\mathcal{O}(\tau^2)$, the approximated average excitation number
\begin{equation}\label{ApproxNumTwoMode}
\begin{aligned}
\bar{n}_{\rm tot}&\approx\frac{\sum_{nm}p_np_m(n+m)\cos^2\Omega_{nm}\tau}{\sum_{nm}(1-g_a^2n\tau^2-g_b^2m\tau^2)e^{-\frac{\omega_a n}{k_BT_a}-\frac{\omega_b m}{k_BT_b}}}\\
&=\frac{\sum_{nm}e^{-\frac{n\omega_a}{k_BT_a}-\frac{m\omega_b}{k_BT_b}}(n+m)\cos^2\Omega_{nm}\tau}{(\bar{n}_{\rm th}+1)(\bar{m}_{\rm th}+1)(1-\Omega^2_{\rm th}\tau^2)},
\end{aligned}
\end{equation}
where $\Omega_{\rm th}\equiv\sqrt{g_a^2\bar{n}_{\rm th}+g_b^2\bar{m}_{\rm th}}$ is defined as the collective thermal Rabi frequency for the double-mode system with $\bar{n}_{\rm th}\equiv\sum_{n}p_nn$ and $\bar{m}_{\rm th}\equiv\sum_{m}p_mm$ representing the average populations of the initial thermal states for mode-$a$ and mode-$b$, respectively. Similarly, a ``singularity'' $\tau_{\rm opt}$ emerges in the same form as Eq.~(\ref{OptIntervalSingle}) with a modified or collective $\Omega_{\rm th}$.

As plotted by the vertical red lines in Figs.~\ref{OptimalInterval}(a) and (b), the reciprocal value of the thermal Rabi frequency apparently determined by the initial temperatures can serve as an approximated expression for the optimized measurement interval. Equation~(\ref{OptIntervalSingle}) is much to our anticipation that a more frequent measurement is required for a higher initial temperature. In a JC-like model, coupling to a higher temperature resonator gives rise to a faster transitions between the ground state and the excited states of the ancillary system and a shorter period for the ground-state population transfers to the excited states. Then a more frequent measurement is demanded to interrupt this undesired process for cooling.

In addition, the preceding derivation that is based on the resonant condition could be generalized to the off-resonant condition. In this case, the denominator in either Eq.~(\ref{OneModeApproxNum}) or Eq.~(\ref{ApproxNumTwoMode}) becomes $1-\Omega_{\rm th}^2\tau^2[\sin(x)/x]^2$ with $x\equiv\delta\tau/2$. Due to the fact that $\lim_{\delta\rightarrow 0}\sin(x)/x=1$, the approximated result in Eq.~(\ref{OptIntervalSingle}) still holds under the condition with a nonvanishing but sufficiently small $\delta$. The leading correction is indeed in the order of $\mathcal(\delta^4)$.

\begin{figure}[htbp]
\centering
\includegraphics[width=0.95\linewidth]{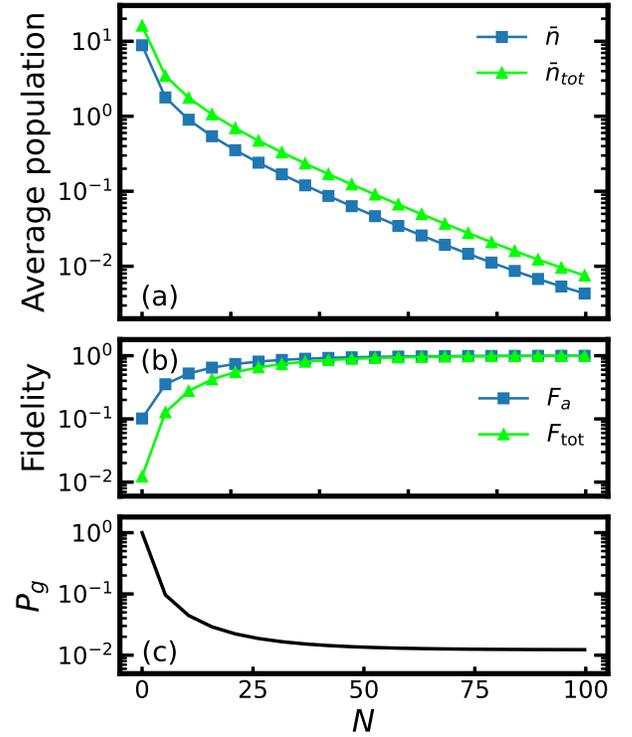}
\caption{Cooling performance under the optimized double-mode-cooling protocol with equal-time-spacing measurements presented by (a) the average population for mode-$a$ $\bar{n}$ and for both modes $\bar{n}_{\rm tot}$, (b) the fidelities of mode-$a$ in its ground state $F_a\equiv\langle n=0|{\rm Tr_b}[\rho_{ab}(N\tau)]|n=0\rangle$ and the overall system in the overall ground state $F_{\rm tot}\equiv\langle n=m=0|\rho_{ab}(N\tau)|n=m=0\rangle$, and (c) the survival probability $P_g(N\tau)$ of detecting the ancillary qutrit in its ground state $|g\rangle$. The parameters are set as $\omega_a=1.4$ GHz, $\omega_b=1.2\omega_a$, $T_a=T_b=0.1$ K, $g_a=g_b=0.04\omega_a$, and $\delta=0.01\omega_a$. }\label{TwoModeCooling}
\end{figure}

Now we consider cooling down two mechanical microresonators in Gigahertz~\cite{MechanicalResonator1,MechanicalResonator2} with a fixed optimized measurement interval. In the numerical simulation, the eigenfrequencies are chosen as $\omega_a=1.4$ GHz, $\omega_b=1.2\omega_a$, and their coupling strengths with the ancillary qutrit are set as $g_a=g_b=0.04\omega_a$. Therefore we have $\bar{n}_{\rm th}=8.85$, $\bar{m}_{\rm th}=7.30$, and $\tau_{\rm opt}=6/\omega_a$ at the initial moment. The other parametric setting is the same as Fig.~\ref{OptimalInterval} for the single measurement. Figure~\ref{TwoModeCooling}(a), (b), and (c) are used to show the performance of our nondeterministic cooling protocol in terms of the average populations, the ground-state fidelities, and the successful probability of detecting the qutrit in its ground state, respectively.

In Fig.~\ref{TwoModeCooling}(a), it is found that the average populations for both mode-$a$ and the overall double modes could be considerably reduced by more than three orders in magnitude through dozens of measurements. In particular, the total average population decreases from $\bar{n}_{\rm tot}\approx16$ to below $\bar{n}_{\rm tot}=0.1$ after $N=50$ measurements and continuously to lower than $\bar{n}_{\rm tot}\approx 8\times10^{-3}$ after $N=100$ measurements. According to the Maxwell-Boltzmann distribution, the average population can be understood by the effective temperature as a direct measure of cooling, which can be defined as
\begin{equation}\label{EffectiveTemperature}
T_{\rm eff}^a=\frac{\omega_a}{k_B\ln(1+1/\bar{n})}, \quad T_{\rm eff}^b=\frac{\omega_b}{k_B\ln(1+1/\bar{m})},
\end{equation}
for mode-$a$ and mode-$b$, respectively. In terms of the effective temperature, mode-$a$ is cooled from $0.1$ K down to $T_{\rm eff}^a\approx2.0$ mK, and mode-$b$ is cooled down to $T_{\rm eff}^b\approx1.9$ mK, demonstrating a reduction about two orders in magnitude.

We can also show the cooling efficiency via the ground-state fidelities of the single mode-$a$ $|n=0\rangle$ and the overall resonator system $|n=m=0\rangle$. In Fig.~\ref{TwoModeCooling}(b), the ground-state fidelity of mode-$a$ $F_a$ is enhanced to $0.95$ after $N=50$ measurement and approaches $0.997$ after $N=100$ measurements. And for the overall system, $F_{\rm tot}$ is over $0.90$ after $N=50$ measurement and approaches $0.995$ after $N=100$ measurements. Figure~\ref{TwoModeCooling}(c) demonstrates the cost for our measurement-based cooling by the successful or survival probability of measurement $P_g(N)$ in Eq.~(\ref{PgN}). It is shown that by repeating periodical measurements, $P_g(N)$ firstly decreases with a great rate. And after about $N=25$ rounds, it gradually approaches an asymptotic value rather than undergoes an exponential decay, which means the ancillary qutrit under measurement becomes fixed in its ground state. This suppression in dynamics simulates the quantum Zeno effect. In our measurement-based cooling protocol, dozens of the projective measurements on qutrit lead it to remain in its ground-state subspace and the successful probability therefore becomes almost invariant with time.

\begin{figure}[htbp]
\centering
\includegraphics[width=0.95\linewidth]{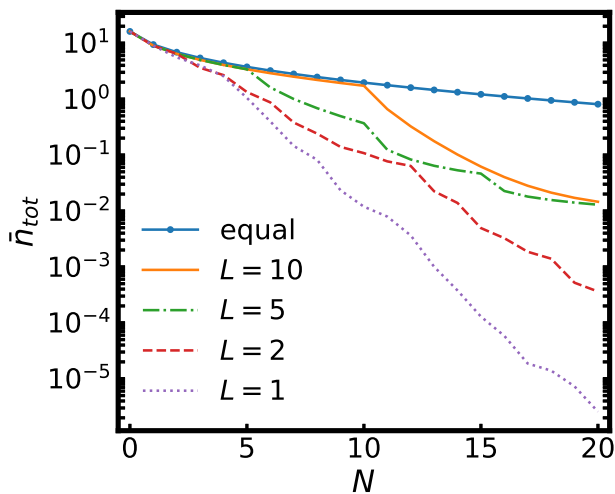}
\caption{The total average population of the double-resonator system under equal-time-spacing and unequal-time-spacing measurement-based cooling strategies. The blue solid line with circle marker represents the strategy with a fixed optimized measurement interval by Eq.~(\ref{OptIntervalSingle}). The orange solid line, the green dot-dashed line, the red dashed line, and the purple dotted line represent the strategies with unequal-time-spacing measurements, where the measurement interval is updated after every $L=10, 5, 2, 1$ rounds of free-evolution and projective measurement, respectively, according to Eq.~(\ref{OptIntervalMulti}). The other parameters are the same as those in Fig.~\ref{TwoModeCooling}. }\label{UnequalTime}
\end{figure}

With a feedback mechanism can be established in the cooling setup, the collective {\em thermal} Rabi frequency in the optimized measurement-interval expression~(\ref{OptIntervalSingle}) could be interpreted as a function of the resonators' average population after the last-round measurement. Then the optimized protocol with equal-time-spacing can be updated to an unequal-time-spacing version by setting
\begin{equation}\label{OptIntervalMulti}
\tau_{\rm opt}\rightarrow\tau_{\rm opt}(t)=\frac{1}{\Omega_{\rm th}(t)}, \quad \Omega_{\rm th}(t)\equiv\sqrt{g_a^2\bar{n}(t)+g_b^2\bar{m}(t)}
\end{equation}
where $\bar{n}(t)\equiv\sum_{n}p_n(t)n$ and $\bar{m}(t)\equiv\sum_{m}p_m(t)m$ represent the {\em current} average populations of mode-$a$ and mode-$b$, respectively. In another word, the sequence of the measurement intervals now becomes $\{\tau_{\rm opt}(t_1), \tau_{\rm opt}(t_2), \cdots, \tau_{\rm opt}(t_N)\}$ with $t_{i>1}=\sum_{j=1}^{j=i-1}\tau_{\rm opt}(t_j)$, instead of a constant $\tau_{\rm opt}$ by the {\em initial} average populations. Note $\tau_{\rm opt}(t_1)=\tau_{\rm opt}$. Clearly we have $\tau_{\rm opt}(t_i)\leq\tau_{\rm opt}(t_{i+1})$, since $p_n(t_1)\leq p_n(t_2)$ and $p_m(t_1)\leq p_m(t_2)$ during the cooling process.

Under this strategy, the overall state of the two resonator modes in Eq.~(\ref{rhoab2}) should be updated to
\begin{equation}\label{rhodifftime}
\begin{aligned}
&\rho_{ab}\left[\sum_{i=1}^N\tau_{\rm opt}(t_N)\right]\\ =&\frac{\sum_{nm}\prod_{i=1}^N|\alpha_{nm}[\tau_{\rm opt}(t_i)]|^2p_np_m|nm\rangle\langle nm|}{P_g(N)}
\end{aligned}
\end{equation}
after $N$ measurements, where the survival/success probability becomes
\begin{equation}\label{PgNdifftime}
P_g(N)=\sum_{nm}\prod_{i=1}^N|\alpha_{nm}[\tau_{\rm opt}(t_i)]|^2p_np_m.
\end{equation}
Note now the cooling coefficient $\alpha_{nm}[\tau_{\rm opt}(t_i)]$ deviates significantly from that in Eq.~(\ref{CoolingCoeffExp}) due to the time-varying argument. Then the populations over any Fock states besides the ground state are no longer under protection as if $N>1$. After a round of free-evolution and postselection by projective measurement, the average population of resonators is certainly reduced. Then a constant measurement-interval $\tau_{\rm opt}$ by the initial temperature in Eq.~(\ref{OptIntervalSingle}) becomes less optimal. Therefore, one can expect a dramatic promotion in cooling performance under the unequal-time-interval strategy.

Regarding the experimental cost in practice, one can update $\tau_{\rm opt}$ with an available rate according to Eq.~(\ref{OptIntervalMulti}). In Fig.~\ref{UnequalTime}, we present the cooling performance by the equal-time-spacing strategy and the unequal-time-spacing strategies under various updating rates. $L=10$ means that the total average population follows the same behavior as the equal-time-spacing strategy in the first $10$ rounds. And then $\tau_{\rm opt}$ is updated according to Eq.~(\ref{OptIntervalMulti}) with $t=10\tau_{\rm opt}$. When $L=1$, it means that the period of each run of evolution-and-measurement has been timely iterated. In comparison of the results in Fig.~\ref{UnequalTime}, the cooling performance finds better improvement with more frequently updating of the optimal interval. A dramatic effect on cooling presents merely by one-time updating that $\bar{n}_{\rm tot}$ is reduced by $3$ orders in magnitude (see the orange solid line). If one can timely iterate the optimal interval according to Eq.~(\ref{OptIntervalMulti}), then the total average population is reduced by $6$ orders in magnitude only by $N=20$ measurements, showing an overwhelming advantage over the equal-time-spacing strategy.

In the absence of the any control, such as the cooling by measurement, the coherence time of a microwave mechanical resonator with a frequency of Gigahertz and a damping rate of $10^4$ Hz~\cite{MechanicalResonator2,OptomechanicsReview} is about $10 \mu$s. Starting from $T\approx 0.1$ K or $\bar{n}\approx 10$, it is found that the full performing time for both the constant-measurement-interval protocol with $N=100$ and the iterative-measurement-interval protocol with $N=10$ is about $0.1 \mu$s. Then our protocols will not be significantly plagued by the nonunitary effects from decoherence, at least during dozens of rounds of evolution-and-measurement.

\section{Multi-mode cooling}\label{MultimodeCoolingSec}

\begin{figure}[htbp]
\centering
\includegraphics[width=0.8\linewidth]{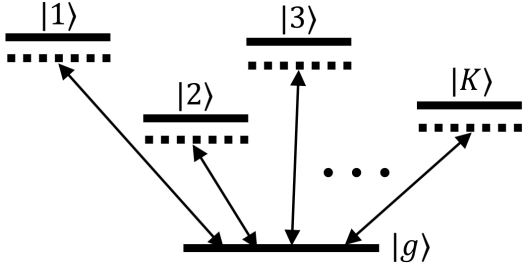}
\caption{Sketch of the ancillary system for the multi-mode cooling protocol, which consists of multiple excited states and a common ground state. The transition between the $k$th excited state and the ground state $|k\rangle\leftrightarrow|g\rangle$ is coupled to the resonator mode-$k$. }\label{MultimodeModel}
\end{figure}

Our simultaneous-cooling protocol as well as the optimized measurement-intervals by Eqs.~(\ref{OptIntervalSingle}) and (\ref{OptIntervalMulti}) can find straightforward scalability in the multi-mode situation, where $K>2$ excited levels in the ancillary system are individually coupled to $K$ resonators. Each mode is initially prepared at its thermal state $\rho_k^{\rm th}$. For the sketch of the ancillary system in Fig.~\ref{MultimodeModel}, the overall Hamiltonian can be written as
\begin{equation}
H=\sum_{k=1}^K\left[\mu_k|k\rangle\langle k|+\nu_kc_k^\dagger c_k+g_k(\sigma_k^+c_k+\sigma_k^-c_k^\dagger)\right],
\end{equation}
where $\mu_k$ and $\nu_k$ are the eigenfrequencies of $k$th excited level $|k\rangle$ of the ancillary system and the $k$th resonator mode, respectively. We have assumed the ground-state energy of the ancillary system to be vanishing. $\sigma_k^-\equiv|g\rangle\langle k|$ and $\sigma_k^+\equiv|k\rangle\langle g|$ are the transition operators of the ancillary system between the level-$k$ and the common ground state $|g\rangle$. $c_k$ and $c_k^\dagger$ are the annihilation and creation operators of the resonator mode-$k$, respectively. $g_k$ labels the particular coupling strength.

\begin{figure}[htbp]
\centering
\includegraphics[width=0.95\linewidth]{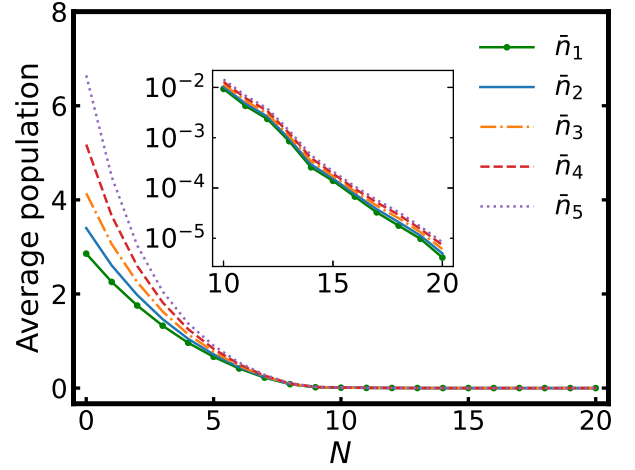}
\caption{Average populations for each mode under the unequal-time-spacing measurement strategy by Eq.~(\ref{OptIntervalMulti}) in a system with $K=5$ resonator modes. The eigenfrequencies of these modes are set as $\nu_k=(0.4+0.2k)\omega_a$, $k=1,2,3,4,5$. These modes are prepared at the same temperature $T=0.05$ K. The other parameters are $\omega_a=1.4$ GHz, $\delta'=0.01\omega_a$, and $g_k=0.04\omega_a$. Inset: a semi-logarithmic scale from $N=10$ to $N=20$. }\label{MultimodeCooling}
\end{figure}

In the rotating frame with respect to $\sum_k\nu_k(|k\rangle\langle k|+c_k^\dagger c_k)$, the Hamiltonian reads
\begin{equation}
H'_I=\sum_{k=1}^{K}\left[\delta_k|k\rangle\langle k|+g_k(\sigma_k^+c_k+\sigma_k^-c_k^\dagger)\right],
\end{equation}
where $\delta_k\equiv\mu_k-\nu_k$ indicates the detuning between the level-$k$ and the mode-$k$. For simplicity, we assume these detunings are on-resonant $\delta_k=\delta'$, $k=1,2,\cdots,K$. Then it is straightforward to find the cooling efficient for arbitrary product Fock state $|n_1,n_2,\cdots,n_K\rangle$ to be
\begin{equation}\label{MultiCoolingCoefficient}
|\alpha_K(\tau)|^2=\frac{\Omega_K^2-\sin^2(\Omega_K\tau)\sum_{k=1}^Kg_k^2n_k}{\Omega_K^2}\leq1,
\end{equation}
where the multi-mode Rabi frequency is defined as
\begin{equation}
\Omega_K\equiv\sqrt{\sum_{k=1}^Kg_k^2n_k+\frac{\delta'^2}{4}}
\end{equation}
with $n_k$ the excitation number of mode-$k$. $|\alpha_K(\tau)|^2=1$ for all the states satisfying $|n_1=n_2=\cdots=n_K=0\rangle$ or $\Omega_K\tau=j\pi$. And the nonunitary evolution operator in Eq.~(\ref{Vg}) for cooling is generalized by $H'_I$ to be
\begin{equation}
\mathcal{V}_g(\tau)=\sum_{n_1,n_2,\cdots,n_K}\alpha_K|n_1,n_2,\cdots,n_K\rangle\langle n_1,n_2,\cdots,n_K|.
\end{equation}
Again the cooling coefficient $|\alpha_K|^2$ ensures the protection over the ground state of all the resonators under $\mathcal{V}_g(\tau)$. Under the unequal-time-spacing strategy for $N$ measurements, the average population for each mode becomes
\begin{equation}\label{barnk}
\bar{n}_k(N)\equiv\langle c_k^{\dagger}c_k\rangle={\rm Tr}\left[c_k^{\dagger}c_k\rho_K(N)\right],
\end{equation}
where
\begin{equation}
\rho_K(N)=\mathcal{V}_g^{(N)}\left(\prod_k^K\rho_k^{\rm th}\right)\mathcal{V}_g^{(N)}/P_K(N)
\end{equation}
with $P_K(N)\equiv{\rm Tr}[\mathcal{V}_g^{(N)}(\prod_k^K\rho_k^{\rm th})\mathcal{V}_g^{(N)}]$ with $\mathcal{V}_g^{(N)}\equiv\prod_{i=1}^N\mathcal{V}_g[\tau_{\rm opt}(t_i)]$. And in the general situation, the time-dependent collective thermal Rabi frequency in Eq.~(\ref{OptIntervalMulti}) reads,
\begin{equation}
\Omega_{\rm th}(t)\equiv\sqrt{\sum_{k=1}^Kg_k^2\bar{n}_k(t)},
\end{equation}
where $\bar{n}_k\equiv\sum_{n_k}p_{n_k}(t)n_k$.

In Fig.~\ref{MultimodeCooling}, we demonstrate the scalability of our cooling protocol in a multi-mode system with $K=5$ resonators. To separate the curves of the average populations $\bar{n}_k$ for each mode, their eigenfrequencies are supposed to be $\nu_k=(0.4+0.2k)\omega_a$, $k=1,2,3,4,5$. Then $n_k$'s are initially in the range of $(2\sim 7)$. It is found that after $N=8$ measurements, all of these five resonators are simultaneously cooled down to $n_k\approx0.1$; and after $N=20$ measurements they are cooled down to $n_k\approx10^{-5}$, resulting in a reduction about six orders in magnitude. In a broader perspective, the results in Fig.~\ref{MultimodeCooling} indicate that a single ancillary system with multiple excited levels and a common ground-state could be used to cool down various resonators within a wide range of frequency.

\section{Discussion}\label{Discussion}

\begin{figure}[htbp]
\centering
\includegraphics[width=0.95\linewidth]{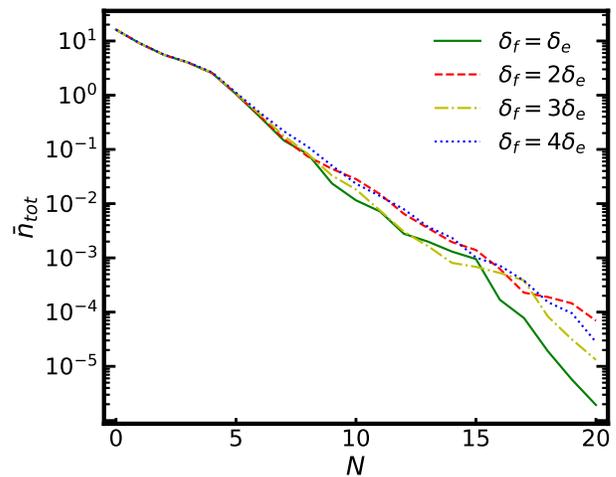}
\caption{Total average population in the double-mode-cooling model under the unequal-time-spacing measurement strategy out of the two-photon resonant condition, i.e., $\delta_e\neq\delta_f$. The detuning between mode-$a$ and level-$e$ of qutrit system $\delta_e$ is fixed as $0.01\omega_a$, and $\delta_f$ between mode-$b$ and level-$f$ varies from $\delta_f=\delta_e$ to $\delta_f=4\delta_e$. The other parameters are $\omega_a=1.4$ GHz, $g_a=g_b=0.04\omega_a$, and $T_a=T_b=0.1$ K.}\label{DifferentDetunings}
\end{figure}

The cooling protocol in Sec.~\ref{HamAndCoolOperater} as well as the cooling coefficient in Eq.~(\ref{CoolingCoeffExp}) is described under the two-photon resonant condition, i.e., $\delta_e=\delta_f$. In Fig.~\ref{DifferentDetunings}, we consider a general situation when these detunings are different from each other. It is found that the total population of the double resonator system $\bar{n}_{\rm tot}$ is almost insensitive to the variation of $\delta_f$ under a fixed $\delta_e$. The distinction between $\delta_f$ and $\delta_e$ only acts a slightly negative effect on the cooling performance, which indicates the robustness of our protocol against the fluctuation in the target-system frequency.

We note the other cooling protocols for multiple resonators have been established in recent literature. The sideband cooling is realized when an energy-damping channel is built to enhance the anti-Stokes scattering~\cite{TwoModeSidebandCooling} or can be improved by breaking the formation of the dark modes decoupled from the full system with introducing a phase-dependent phonon-exchange interaction~\cite{DarkModeBreakingCooling}. The EIT cooling employs an extra ground state in a three-level $\Lambda$-setup to create two independent EIT-structures to suppress the heating process~\cite{EITCoolingTwoMode}. Both of them share a common idea that a decay pathway or a cooling channel is constructed to extract the resonator energy. In sharp contrast, the measurement-based cooling protocols describe a probabilistic process by repeatedly projecting the whole system into the ground state and ignoring the distribution of the ensemble over the high-energy manifolds. Cooling by measurement is essentially a purification process via post-selections, by which the high-energy distributions of resonators are discarded and only the ground state is collected~\cite{ExpOneModeCooling}.

The distinction between the existing cooling protocols for multiple resonators and ours can be also demonstrated by the cooling performance in terms of the average population. For the optomechanical system under the sideband cooling~\cite{TwoModeSidebandCooling}, two nearly degenerate resonators can be cooled from $\bar{n}=40$ down to $\bar{n}\approx2$. For the trapped ion crystal under the EIT cooling~\cite{EITCoolingTwoMode}, it can be cooled from $\bar{n}=6$ down to $\bar{n}\approx0.06$. While our protocols with a constant measurement-interval illustrated in Fig.~\ref{TwoModeCooling} and with an iterative measurement-interval in Fig.~\ref{UnequalTime} can reduce $\bar{n}_{\rm tot}$ by $3$ and $6$ orders in magnitude with a dozen of measurements, respectively.

From an even broader perspective, our cooling-by-measurement protocol finds analogy with the many-atom-state preparation by the photon-signal herald~\cite{Subradiance,DirectedSpontaneousEmission}. Under the phase-match condition in the JC-like interaction, the detection result of the emitted photon heralds the state of interest, such as the single-excitation superradiant and subradiant states and the timed-Dicke state. Clearly the count event or no-count event of photon absorption in those atomic-state-preparation methods plays the same role as the postselection or indirect measurement as in our protocol. The measurements are performed on the ancillary system at a proper time and repeated until the desired outcome is obtained.

\section{Conclusion}\label{Conclusion}

In summary, we present a simultaneous cooling-by-measurement protocol by coupling an ancillary $V$-type three-level system to two nondegenerate target resonators. Analytically, we obtain for the first time an optimized expression $\tau_{\rm opt}\approx 1/\Omega_{\rm th}$ of the measurement interval, which is found to be determined by the average population of the target system before the measurement and the coupling strengths between the target system and the ancillary system. And by iterating $\Omega_{\rm th}$ with the time-varying population, the cooling performance could be greatly improved. Under the unequal-time-spacing measurement strategy, the average population of the resonator-system can be suppressed by $6$ orders in magnitude via only a few dozens of projective measurements on the ground state of the ancillary system. An extra important criterion met by our simultaneous-cooling protocol as well as the optimized measurement-interval is their scaling in the multiple-resonator system. Beyond the resonant condition, our protocol adapts to a wide range of resonator-frequency. Therefore it allows a collective cooling with a high efficiency for arbitrary quantum resonator systems and offers an appealing application for exploring multiphoton process.

\section*{Acknowledgments}

We acknowledge grant support from the National Science Foundation of China (Grants No. 11974311 and No. U1801661).

\bibliographystyle{apsrevlong}
\bibliography{ref}

\end{document}